\documentclass[aip,rsi,reprint]{revtex4-1}
\pdfoutput=1
\usepackage{graphicx}
\usepackage{fullpage}
\usepackage{upgreek}
\usepackage{verbatim}

\begin{document}

\title{An apparatus to manipulate and identify individual Ba ions from bulk liquid Xe}

\author{K.~Twelker}
\affiliation{Physics Department, Stanford University, Stanford CA, USA}
\author{S.~Kravitz}
\affiliation{Physics Department, Stanford University, Stanford CA, USA}
\author{M.~Montero~D\'iez}\thanks{Now at KLA-Tencor, Milpitas CA, USA}
\affiliation{Physics Department, Stanford University, Stanford CA, USA}
\author{G.~Gratta}
\affiliation{Physics Department, Stanford University, Stanford CA, USA}
\author{W.~Fairbank Jr.}
\affiliation{Physics Department, Colorado State University, Fort Collins CO, USA}
\author{J.B.~Albert}
\affiliation{Physics~Department~and~CEEM,~Indiana~University,~Bloomington~IN,~USA}
\author{D.J.~Auty}
\affiliation{Department of Physics and Astronomy, University of Alabama, Tuscaloosa AL, USA}
\author{P.S.~Barbeau}
\affiliation{Department of Physics, Duke University, and Triangle Universities Nuclear Laboratory (TUNL), Durham North Carolina, USA}
\author{D.~Beck}
\affiliation{Physics Department, University of Illinois, Urbana-Champaign IL, USA}
\author{C.~Benitez-Medina}\thanks{Now at Intel, Hillsboro, OR, USA}
\affiliation{Physics Department, Colorado State University, Fort Collins CO, USA}
\author{M.~Breidenbach}
\affiliation{SLAC National Accelerator Laboratory, Stanford CA, USA}
\author{T.~Brunner}
\affiliation{Physics Department, Stanford University, Stanford CA, USA}
\author{G.F.~Cao}
\affiliation{Institute of High Energy Physics, Beijing, China}
\author{C.~Chambers}
\affiliation{Physics Department, Colorado State University, Fort Collins CO, USA}
\author{B.~Cleveland}\thanks{Also SNOLAB, Sudbury ON, Canada}
\affiliation{Department of Physics, Laurentian University, Sudbury ON, Canada}
\author{M.~Coon}
\affiliation{Physics Department, University of Illinois, Urbana-Champaign IL, USA}
\author{A.~Craycraft}
\affiliation{Physics Department, Colorado State University, Fort Collins CO, USA}
\author{T.~Daniels}
\affiliation{Physics Department, University of Massachusetts, Amherst MA, USA}
\author{S.J.~Daugherty}
\affiliation{Physics~Department~and~CEEM,~Indiana~University,~Bloomington~IN,~USA}
\author{C.G.~Davis}\thanks{Now at US Naval Research Lab, Washington DC, USA}
\affiliation{Physics Department, University of Maryland, College Park MD, USA}
\author{R.~DeVoe}
\affiliation{Physics Department, Stanford University, Stanford CA, USA}
\author{S.~Delaquis}
\affiliation{LHEP, Albert Einstein Center, University of Bern, Bern, Switzerland}
\author{T.~Didberidze}
\affiliation{Department of Physics and Astronomy, University of Alabama, Tuscaloosa AL, USA}
\author{J.~Dilling}
\affiliation{TRIUMF, Vancouver BC, Canada}
\author{M.J.~Dolinski}
\affiliation{Department of Physics, Drexel University, Philadelphia PA, USA}
\author{M.~Dunford}
\affiliation{Physics Department, Carleton University, Ottawa ON, Canada}
\author{L.~Fabris}
\affiliation{Oak Ridge National Laboratory, Oak Ridge TN, USA}
\author{J.~Farine}
\affiliation{Department of Physics, Laurentian University, Sudbury ON, Canada}
\author{W.~Feldmeier}
\affiliation{Technische Universitat Munchen, Physikdepartment and Excellence Cluster Universe, Garching, Germany}
\author{P.~Fierlinger}
\affiliation{Technische Universitat Munchen, Physikdepartment and Excellence Cluster Universe, Garching, Germany}
\author{D.~Fudenberg}
\affiliation{Physics Department, Stanford University, Stanford CA, USA}
\author{G.~Giroux}\thanks{Now at Queen’s University, Kingston ON, Canada}
\affiliation{LHEP, Albert Einstein Center, University of Bern, Bern, Switzerland}
\author{R.~Gornea}
\affiliation{LHEP, Albert Einstein Center, University of Bern, Bern, Switzerland}
\author{K.~Graham}
\affiliation{Physics Department, Carleton University, Ottawa ON, Canada}
\author{C.~Hall}
\affiliation{Physics Department, University of Maryland, College Park MD, USA}
\author{M.~Heffner}
\affiliation{Lawrence Livermore National Laboratory, Livermore CA, USA}
\author{S.~Herrin}\thanks{Now at 23andMe, Mountain View, CA, USA}
\affiliation{SLAC National Accelerator Laboratory, Stanford CA, USA}
\author{M.~Hughes}
\affiliation{Department of Physics and Astronomy, University of Alabama, Tuscaloosa AL, USA}
\author{X.S.~Jiang}
\affiliation{Institute of High Energy Physics, Beijing, China}
\author{T.N.~Johnson}
\affiliation{Physics~Department~and~CEEM,~Indiana~University,~Bloomington~IN,~USA}
\author{S.~Johnston}
\affiliation{Physics Department, University of Massachusetts, Amherst MA, USA}
\author{A.~Karelin}
\affiliation{Institute for Theoretical and Experimental Physics, Moscow, Russia}
\author{L.J.~Kaufman}
\affiliation{Physics~Department~and~CEEM,~Indiana~University,~Bloomington~IN,~USA}
\author{R.~Killick}
\affiliation{Physics Department, Carleton University, Ottawa ON, Canada}
\author{T.~Koffas}
\affiliation{Physics Department, Carleton University, Ottawa ON, Canada}
\author{R.~Kr\"ucken}
\affiliation{TRIUMF, Vancouver BC, Canada}
\author{A.~Kuchenkov}
\affiliation{Institute for Theoretical and Experimental Physics, Moscow, Russia}
\author{K.S.~Kumar}
\affiliation{Physics Department, University of Massachusetts, Amherst MA, USA}
\author{D.S.~Leonard}
\affiliation{Department of Physics, University of Seoul, Seoul, Korea}
\author{F.~Leonard}
\affiliation{Physics Department, Carleton University, Ottawa ON, Canada}
\author{C.~Licciardi}
\affiliation{Physics Department, Carleton University, Ottawa ON, Canada}
\author{Y.H.~Lin}
\affiliation{Department of Physics, Drexel University, Philadelphia PA, USA}
\author{R.~MacLellan}
\affiliation{SLAC National Accelerator Laboratory, Stanford CA, USA}
\author{M.G.~Marino}
\affiliation{Technische Universitat Munchen, Physikdepartment and Excellence Cluster Universe, Garching, Germany}
\author{B.~Mong}
\affiliation{Department of Physics, Laurentian University, Sudbury ON, Canada}
\author{D.~Moore}
\affiliation{Physics Department, Stanford University, Stanford CA, USA}
\author{A.~Odian}
\affiliation{SLAC National Accelerator Laboratory, Stanford CA, USA}
\author{I.~Ostrovskiy}
\affiliation{Physics Department, Stanford University, Stanford CA, USA}
\author{C.~Ouellet}
\affiliation{Physics Department, Carleton University, Ottawa ON, Canada}
\author{A.~Piepke}
\affiliation{Department of Physics and Astronomy, University of Alabama, Tuscaloosa AL, USA}
\author{A.~Pocar}
\affiliation{Physics Department, University of Massachusetts, Amherst MA, USA}
\author{F.~Retiere}
\affiliation{TRIUMF, Vancouver BC, Canada}
\author{P.C.~Rowson}
\affiliation{SLAC National Accelerator Laboratory, Stanford CA, USA}
\author{M.P.~Rozo}
\affiliation{Physics Department, Carleton University, Ottawa ON, Canada}
\author{A.~Schubert}
\affiliation{Physics Department, Stanford University, Stanford CA, USA}
\author{D.~Sinclair}
\affiliation{TRIUMF, Vancouver BC, Canada}
\affiliation{Physics Department, Carleton University, Ottawa ON, Canada}
\author{E.~Smith}
\affiliation{Department of Physics, Drexel University, Philadelphia PA, USA}
\author{V.~Stekhanov}
\affiliation{Institute for Theoretical and Experimental Physics, Moscow, Russia}
\author{M.~Tarka}
\affiliation{Physics Department, University of Illinois, Urbana-Champaign IL, USA}
\author{T.~Tolba}
\affiliation{LHEP, Albert Einstein Center, University of Bern, Bern, Switzerland}
\author{D.~Tosi}\thanks{Now at University of Wisconsin, Madison, WI, USA}
\affiliation{Physics Department, Stanford University, Stanford CA, USA}
\author{J.-L.~Vuilleumier}
\affiliation{LHEP, Albert Einstein Center, University of Bern, Bern, Switzerland}
\author{J.~Walton}
\affiliation{Physics Department, University of Illinois, Urbana-Champaign IL, USA}
\author{T.~Walton}
\affiliation{Physics Department, Colorado State University, Fort Collins CO, USA}
\author{M.~Weber}
\affiliation{Physics Department, Stanford University, Stanford CA, USA}
\author{L.J.~Wen}
\affiliation{Institute of High Energy Physics, Beijing, China}
\author{U.~Wichoski}
\affiliation{Department of Physics, Laurentian University, Sudbury ON, Canada}
\author{L.~Yang}
\affiliation{Physics Department, University of Illinois, Urbana-Champaign IL, USA}
\author{Y.-R.~Yen}
\affiliation{Department of Physics, Drexel University, Philadelphia PA, USA}
\author{Y.B.~Zhao}
\affiliation{Institute of High Energy Physics, Beijing, China}

\date{\today}

\begin{abstract}
We describe a system to transport and identify barium ions produced in liquid xenon, as part of R\&D towards the second phase of a double beta decay experiment, nEXO.  The goal is to identify the Ba ion resulting from an extremely rare nuclear decay of the isotope $^{136}$Xe, hence providing a confirmation of the occurrence of the decay.  This is achieved through Resonance Ionization Spectroscopy (RIS). In the test setup described here, Ba ions can be produced in liquid xenon or vacuum and collected on a clean substrate. This substrate is then removed to an analysis chamber under vacuum, where laser-induced thermal desorption and RIS are used with time-of-flight (TOF) mass spectroscopy for positive identification of the barium decay product.
\end{abstract}


\maketitle 

\section{Introduction}
\label{sec:Introduction}
The neutrino is the only fundamental particle that is potentially a Majorana fermion, a property that could be revealed in neutrinoless double-beta decay ($0\nu\beta\beta$)~\cite{Avignone:2008eo}.  $^{136}$Xe is a candidate nucleus for this process with a half-life in excess of  $10^{25}$~years~\cite{Albert:2014ch,Gando:2013wx}.  nEXO will search for $0\nu\beta\beta$ in a liquid xenon Time Projection Chamber (TPC)~\cite{Fancher:1979wf} with 5 tons of xenon enriched to $\sim 90$\% in $^{136}$Xe. 

In a possible future upgrade of nEXO, most backgrounds obscuring the $0\nu\beta\beta$ signal could be eliminated through the identification of $^{136}$Ba (``Ba tagging''), the daughter nucleus of the decay.  The majority of background in experiments like nEXO comes from radioactive decays of impurities in the materials used in the detector construction, despite great efforts to use radiopure materials~\cite{Auger:2012up}.  Because such decays do not produce Ba, identification of the daughter for each candidate $0\nu\beta\beta$ event would eliminate these backgrounds~\cite{moe_detection_1991}.  Backgrounds from natural Ba are expected to be very low within the detector, particularly since only a small volume of Xe around the decay is relevant.   The rate of false positives will be measured in real-time by directing the Ba tagging system to sample random locations in the LXe.  The sensitivity of nEXO without Ba tagging is expected to be ${T_{1/2}^{0\nu\beta\beta}<6.0\times10^{27}}$~years. The addition of Ba tagging could extend the sensitivity to ${T_{1/2}^{0\nu\beta\beta}<3.2\times10^{28}}$~years.  This is the only proposed concept of a technique that can be projected to reach part of the so-called ``normal hierarchy'' of neutrino masses.  This paper describes an apparatus built to test removal and identification of Ba from liquid xenon (LXe).  The goal is to develop this Ba tagging technology toward unit efficiency so that a properly engineered system can be installed in a future version of the nEXO detector.

Our approach is to electrostatically drift Ba$^+$ ions to a clean substrate, adsorb them, then transport the substrate to vacuum with the Ba attached.   Neutral Ba atoms are then removed from the substrate using laser induced thermal desorption resulting in individual atoms in vacuum, making their atomic spectroscopy available for identification.  Resonance Ionization Spectroscopy (RIS) is used to ionize Ba from the desorbed plume of neutral atoms.  This highly selective process allows Ba to be identified despite the possible presence of other elements.  RIS can also be very efficient~\cite{hurst_resonance_1979}, making the detection of single Ba atoms possible. 

In order to model and test the final application, the development of this process must use very small numbers of Ba atoms.  Removal of Ba from the substrate leads to elimination of the signal, making this measurement intrinsically destructive.  Therefore careful control over all aspects of the system is necessary for effective operation.

To help identify the signal, a time-of-flight (TOF) spectrometer separates Ba ions from other species that are ionized due to various undesirable effects.  This extra selectivity complements and cross-checks that of the RIS process.  In order to maximize the recovery of Ba, the TOF spectrometer is optimized for ion transport efficiency.  The TOF of each ion detected is recorded for later analysis.  A practical realization of the Ba tagging technique in double-beta decay may use a TOF system similar to the one described here or, for optimal signal identification, an ion trap with spectroscopic readout such as the one described in~\cite{Green:2007vq}.

In the system, Ba can be produced by two sources using radioactive decays; one is driven by the recoils of an $\alpha$~decay~\cite{montero_diez_simple_2010} (appropriate for initial tests in vacuum), while the other is a $^{252}$Cf fission source that produces Ba ions directly (to be used for tests in liquid xenon), similar to radioactive beam sources~\cite{savard_radioactive_2008}.  Because these sources produce Ba ions at a fixed location, there is no need to search the entire LXe volume.  Mechanical systems to position a probe at any location within a large LXe TPC will be developed separately for nEXO.
\begin{figure*}
	\includegraphics[width=6.4in, bb=0 0 457 234]{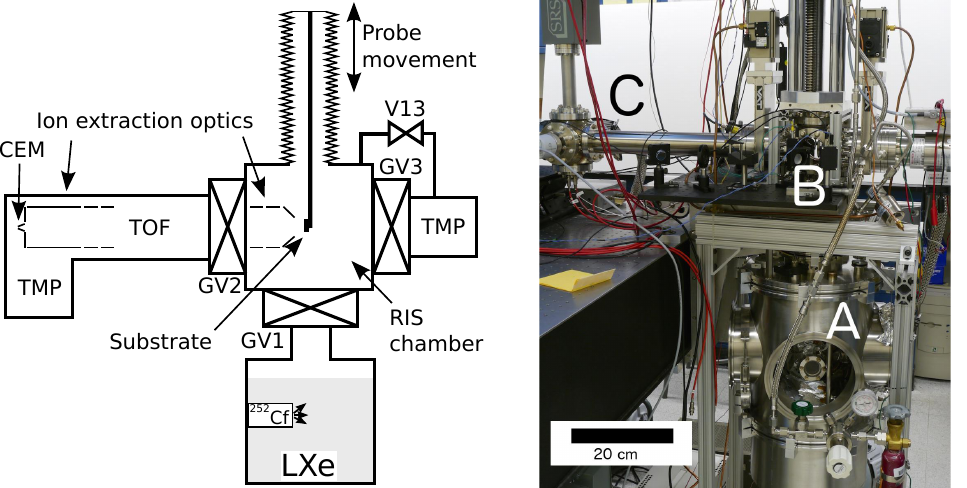}
	\caption{Functional diagram (left) and photograph (right) of the Ba tagging system.  In the photograph {\bf A} is the LXe cell, {\bf B} is the RIS chamber, and {\bf C} is the TOF spectrometer.  Much of the Xe handling system is omitted for simplicity (see Fig.~\ref{fig:XenonPlumbing}).  \label{fig:OperationSchematic}}
\end{figure*}

We describe a system, shown in Figure~\ref{fig:OperationSchematic}, capable of testing extraction and identification of Ba$^+$ ions intentionally deposited in LXe.  The substrate mounted on the end of the probe is moved into the LXe to deposit Ba ions from the source.  The substrate is then moved to the RIS chamber, and the gate valve (GV1) is closed to isolate the RIS chamber from the LXe cell.  The RIS chamber is evacuated by a combination of cryopumping and the turbomolecular pump (TMP) through GV3, and finally GV2 is opened to the TOF spectrometer for identification of atoms desorbed and re-ionized.

The LXe conditions are similar to those of a LXe TPC: $\sim 900$~Torr pressure at a temperature of $165$~K.  The apparatus includes systems to cool and condense LXe, and can test Ba tagging from Ba production through retrieval and identification.
		
\section{Liquid xenon cell and probe}
\label{sec:LXeVacuumJacket}
This system condenses xenon in a $1$~liter copper cell cooled using liquid nitrogen (LN$_2$).  The cell is insulated from room temperature by a vacuum of $10^{-5}$~Torr.  Radiative heating across the vacuum is limited by $10$~layers of super insulation.  
\begin{figure}
	\includegraphics[width=3.0in, bb=0 0 216 145]{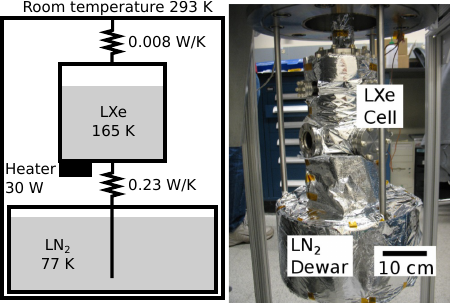}
	\caption{Thermal model of the LXe cell and LN$_2$ reservoir (left) and a photograph (right).  The LXe cell is cooled by LN$_2$ held in a dewar mounted inside the same vacuum insulation.   The photograph shows the LN$_2$ dewar and LXe cell wrapped in super insulation. \label{fig:ThermalDiagram}}
\end{figure} 

The 10~liter LN$_2$ dewar is mounted in the same vacuum  as the LXe cell, as shown in Figure~\ref{fig:ThermalDiagram}.  The cell is connected to the LN$_2$ bath through a copper heat-transfer strap.  Because the thermal conductivity of this strap is much greater than the thermal conductivity of the tube leading to room temperature, the equilibrium temperature is below the temperature required to maintain the Xe in liquid phase ($165$~K), typically requiring about $20$~W of heating power.  The heaters are controlled through a PID loop using the cell temperature measured by Resistive Thermal Devices (RTDs) read out by a PLC system\cite{fn1}.  To ensure thermal stability and uniformity, the LXe cell is built with $1.9$~cm thick copper walls.

Access to the cell is provided through several CF flanges brazed into the copper cell body.  The cell is supported from the top CF 6.00'' flange, through which the probe passes.   Eight CF flanges ($4\times$~CF 2.75'', $4\times$~CF 1.33'') are available to mount viewports, sources, and voltage feedthroughs.  The Ba source is mounted on one of these CF 2.75'' ports, while the other three are reserved for optical access to the LXe.

\subsection{Xenon delivery and recovery}
\label{sub:XePlumbing}
\begin{figure*}
	\includegraphics[scale=1, bb=0 0 476 239]{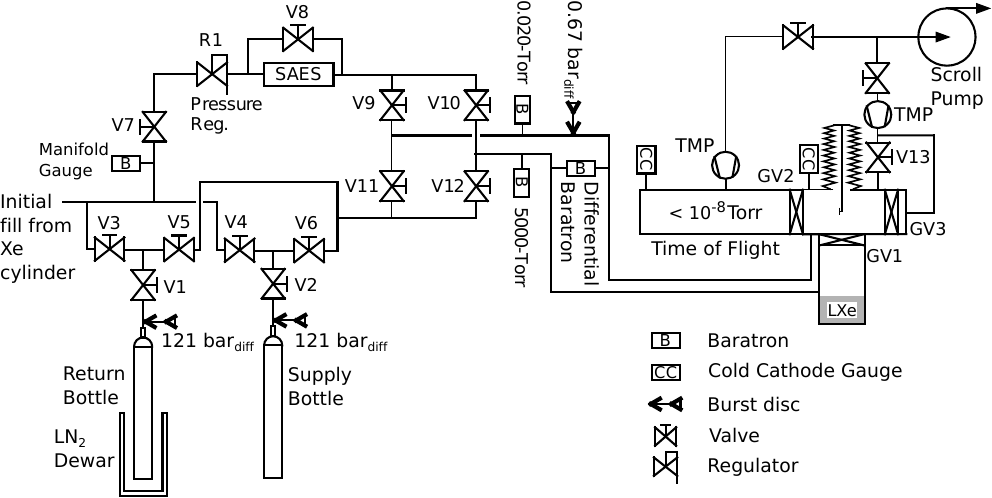}
	\caption{Vacuum and xenon systems.  Valves are configurable to feed xenon from either bottle, and recover it to the other, immersed in LN$_2$.  Valves allow feeding and recovering LXe from both the LXe cell and the RIS chamber, which allows sampling the substrate in vacuum while maintaining the LXe in the cell.  Turbomolecular pumps (TMP) maintain the TOF at UHV and are used to bring the RIS chamber to vacuum.  \label{fig:XenonPlumbing}}
\end{figure*}

A schematic view of the vacuum and Xe systems is shown in Figure~\ref{fig:XenonPlumbing}.  Since the measurement requires shuttling the probe between the LXe ($\sim 900$~Torr) and the TOF spectrometer (vacuum), the RIS chamber functions as a load-lock isolated by three gate valves (Fig.~\ref{fig:OperationSchematic}).  The LXe level must be maintained while GV1 is closed, which requires xenon feed to and recovery from the cell.  At the same time, xenon must be added to or removed from the RIS chamber either to pump the RIS chamber to vacuum or to balance the pressure across GV1 before opening it.  Xenon feed to and cryopumping from both the cell and the RIS chamber are controlled by pneumatically-actuated valves (V9-V12).  

Xenon is supplied to the system from a two-bottle manifold: one serves as the supply, the other serves as a recovery/cryopump bottle.  For any given run, the manual valves are configured to feed xenon from the supply bottle through the regulator and purifier\cite{fn2} to fill the cell.  The recovery bottle is empty at the beginning of a measurement cycle and is immersed in LN$_2$.  At the end of a run all of the xenon is transferred to the recovery bottle.  The recovery bottle is then warmed up and serves as the supply bottle for the next run.   Bottle manifold valves (V1-V6), V7 and V8  are manual. 

Capacitance manometers measure pressure throughout the xenon system.  Pressure in the LXe cell is measured by a $5000$-Torr full-scale gauge, while pressure in the RIS chamber is measured by a $20$-mTorr full-scale gauge\cite{fn3}.  Pressure across GV1 is measured by a differential Baratron\cite{fn4}.  Vacuum is measured with cold cathode gauges\cite{fn5} on the RIS chamber and the TOF spectrometer.  

All temperatures and pressures are recorded at a $1$~Hz rate and serve as input to a PID loop that controls valve states and the two 30~W resistors that are used to heat the LXe cell.  Pneumatic valves (V9-V13 in Figure~\ref{fig:XenonPlumbing}) are controlled by a digital output to add or remove xenon from the system.  The gate valves (GV1-GV3 in Figures~\ref{fig:OperationSchematic}~and~\ref{fig:XenonPlumbing}) are also controlled by the PLC system. 

The Ba collection cycle begins with all three gate valves (GV1-3) closed, with the RIS chamber under vacuum.  GV1 can only be opened after the Xe pressure across it has been balanced by filling the RIS chamber through V9.  The probe can then move the substrate into the LXe.  Once Ba has been deposited and the substrate has been retracted to the RIS chamber, the system is configured to recover the Ba: GV1 is closed, the RIS chamber pumped to vacuum.  Cryopumping to the recovery bottle reduces the pressure in the RIS chamber from $\sim 900$~Torr to less than $1$~Torr in $\sim 1$~minute.  A high-impedance bypass valve (V13) opens to bring the pressure to below $10^{-3}$ Torr with the turbomolecular pump always running at full speed.  GV3 then opens when the RIS chamber is at $10^{-3}$~Torr and finally GV2 opens with the RIS chamber pressure below $10^{-5}$~Torr.  This procedure loses roughly 0.02~g of Xe by pumping it into the atmosphere. 

\section{Barium ion sources}
\label{sec:Sources}
Barium ion sources are necessary to produce the small amount of Ba needed to develop this technique. For some tests the Ba has to be produced in LXe, adding special constraints because of its high density.  Sources producing large and hard to control quantities of Ba are not considered. 

Currently, a Gd-driven Ba ion source is used in vacuum and the use of a fission source is planned for LXe.  In the Gd-driven source recoiling nuclei from the $\alpha$ decay of $^{148}$Gd sputter Ba and BaF atoms and ions from the BaF$_2$ layer.  Most atoms are emitted with energies under $1$~keV, but the energy distribution has a long tail extending to $89$~keV.  This source has been described in a separate publication~\cite{montero_diez_simple_2010}.

Because the Gd-driven source produces Ba ions with a very short range in LXe we plan to produce Ba ions in LXe using a $^{252}$Cf source\cite{fn6}, electroplated on Pt in such a way as to allow the fission fragments to escape.  Each fission produces two neutron-rich isotopes, which $\beta$~decay to stable isotopes.  A few percent of the fission products and subsequent decay chains of $^{252}$Cf produce Ba isotopes. The energy of fission fragments averages around $70$~MeV, enough to deliver the Ba $\sim30$~$\upmu$m into the LXe, calculated using SRIM~\cite{Ziegler:2010wt}.  

The accumulation of Ba in the LXe from a $1$~kBq source is calculated from the known fission yields~\cite{England:C6UzhYbB} and shown in Table~\ref{tab:BaIsotopes}. This analysis assumes that there are no fission products in the LXe at the start of deposition.  Deposition begins when the substrate is moved in front of the source and is biased to attract ions.  The source will be biased to readsorb ions resulting from fissions while the substrate is not positioned in front of the source.  
\begin{table}
	\begin{center}
	\begin{tabular}{ccc}
		\hline\hline
		Isotope&Half-life&Accumulation after 30~min\\
		\hline
		$^{138}$Ba& Stable&375\\
		$^{139}$Ba& 83.1 m&1760\\
		$^{140}$Ba& 12.8 d&3130\\
		$^{141}$Ba& 18.3 m&1950\\
		$^{142}$Ba& 10.6 m&1360\\
		\hline\hline
	\end{tabular}
	\end{center} 
	\caption{Isotopes of Ba emitted in large numbers from a $^{252}$Cf source. }
	\label{tab:BaIsotopes}
\end{table}

The Ba isotopes are primarily produced as ions by $\beta$ decays from Cs isotopes.  $^{134}$Ba through $^{137}$Ba are not produced in large numbers because the corresponding Cs isotopes are long lived.  Heavier isotopes ($^{143}$Ba and heavier) are too short-lived to be of use in this system~\cite{nndc,England:C6UzhYbB}.  For $^{138}$Ba, the only stable isotope produced in a significant amount, the source will produce 375 ions in the first $30$~minutes.  Other isotopes, such as $^{139}$Ba through $^{142}$Ba, are relatively long-lived, and will be detectable in this system.  Detection of these isotopes will help distinguish between Ba intentionally deposited and any remaining naturally existing background.

\section{Laser Setup}
\label{sec:Lasers}
Laser induced thermal desorption (LITD) is a convenient method for removing a few atoms from a  substrate~\cite{Burgess:1986gy}.  Barium is then selectively ionized from the desorbed plume of neutral atoms using RIS. In order to allow the desorbed plume of atoms to move away from the substrate and mitigate image-charge effects, the RIS lasers are delayed by $1~{\rm \upmu s}$ with respect to the LITD pulse. 

In addition to desorbing neutral atoms, LITD can directly produce ions which are a background to the Ba signal.  In order to characterize the various backgrounds during data collection, the LITD and RIS laser pulses are cyclically suppressed.  The following three states are alternated:
\begin{itemize}
\item Both LITD and RIS pulses are retained.  Ba is ionized selectively, along with possible backgrounds due to the LITD laser alone.
\item Only the LITD pulses are retained.  Ions directly generated by the desorption laser alone can be identified and subtracted from the data in the first cycle.
\item Only the RIS pulses are retained.  This configuration can be used  to measure the  background that may be produced if the RIS pulses are not properly shaped in space and desorb ions from the substrate.  
\end{itemize}
All lasers fire at a rate of $10$~Hz and a system of mechanical shutters implements the pulse sequence described above. This arrangement provides better stability in the pulse energy as the lasers operate in a thermal steady state.  

\begin{figure}
	\includegraphics[scale=1, bb=0 0 236 350]{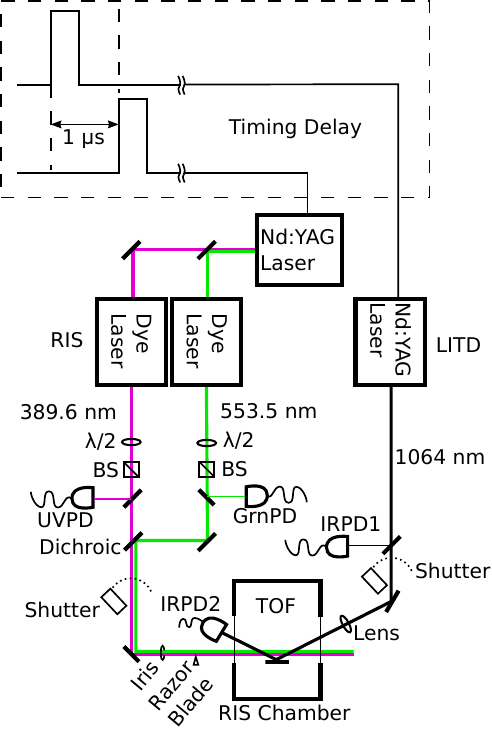}
	\caption{(Color online) Laser optics and timing.   \label{fig:LaserOptics}}
\end{figure}

Figure~\ref{fig:LaserOptics} provides a simplified schematic of the optical setup.  The pulse energy of the desorption laser ($1064$~nm) is measured by two infrared photodiodes (IRPD1 and IRPD2 before and after hitting the substrate, respectively).  Dye lasers pumped by a second Nd:YAG provide the RIS pulses.  Both ionization lasers use energy control through $\lambda /2$ waveplates and polarizing beam splitting cubes (BS), and each is measured by a photodiode (GrnPD and UVPD).   

\subsection{Laser Induced Thermal Desorption}
\label{sub:Desorption}
The Ba atoms must be desorbed in a plume that can be ionized using pulsed dye lasers.  We use a pulsed $1064$~nm Nd:YAG laser\cite{fn7} to thermally desorb atoms from the substrate.  

The Nd:YAG laser produces a Gaussian beam profile, which is loosely focused onto the substrate using a $200$~mm focal length lens.  Because of the 70~degree angle of incidence, this results in an elliptical Gaussian profile measuring $275~\upmu{\rm m}\times 750~\upmu{\rm m}$ (Gaussian $\sigma$) on the substrate. The LITD intensity is typically 2~MW/cm$^2$ at the center of the spot averaged over the 7~ns laser pulse. 

The desorption laser pulse energy is carefully controlled to limit ionization from the desorption process.  Individual pulses of the desorption laser are measured by two separate photodiodes\cite{fn8}.  The first infrared photodiode (IRPD1 in Fig.~\ref{fig:LaserOptics}) measures each individual IR pulse before the beam enters the RIS chamber.  Measurement of the reflected beam off of the substrate (IRPD2) is used to aid in positioning the laser spot on the substrate.  Both diode readings are recorded for each laser pulse and are calibrated to a powermeter\cite{fn9}.

The desorption laser power must be set to minimize backgrounds but ensure desorption of neutral Ba from the surface.  The desorption laser pulse energy is increased until the RIS Ba peak appears indicating that Ba has been desorbed as a neutral atom and resonantly ionized.  If the desorption laser produces significant backgrounds at the TOF of Ba, then the pulse energy is reduced.

\subsection{Resonance Ionization Spectroscopy lasers}
\label{sub:RIS}
\begin{figure}
	\includegraphics[scale=1,bb=0 1 234 193]{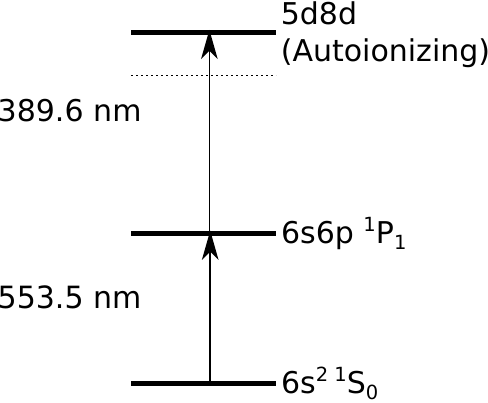}
	\caption{The RIS scheme.  The dashed line indicates the approximate energy of the first ionization threshold.  \label{fig:RISScheme}}
\end{figure}

Tunable dye lasers are used for the RIS process. These lasers are pumped by the second and third harmonics of a separate Nd:YAG laser\cite{fn10}.  The RIS scheme, shown in Figure~\ref{fig:RISScheme}, uses two wavelengths\cite{Willke:1993wt,Camus:1983vv}: $553.5$~nm\cite{fn11} and $389.6$~nm\cite{fn12}.  Both wavelengths are periodically verified using a wavemeter\cite{fn13} to ensure stability.  

The RIS process requires that both RIS beams are spatially and temporally overlapped.  Spatial overlap is achieved by using a dichroic mirror that transmits UV light and reflects green light, as shown in Figure~\ref{fig:LaserOptics}.  A razor blade clips the beam profile and an iris blocks the tails of the Gaussian beam obtaining a $5$~mm-diameter semicircular beam profile that optimally overlaps the desorbed plume of atoms and grazes the substrate without striking it.  This profile overlaps most desorbed atoms from the typical $4~{\rm mm}\times4~{\rm mm}$~scan of the LITD laser. 

Tracking the energies of the laser pulses is crucial to ensure efficient RIS. In particular, the pulse energies of the dye lasers decrease in intensity over time as the dye is bleached.  Each laser uses a separate photodiode pickoff to provide a relative energy measurement on a pulse-by-pulse basis. A separate cali\-bration provides the pulse energy at the window to the vacuum chamber.  All photodiodes are read out by flash ADCs\cite{fn14}. 

\section{Time of flight spectrometer}
\label{sec:TOF}
\begin{figure*}
\onecolumngrid
	\includegraphics[scale=1]{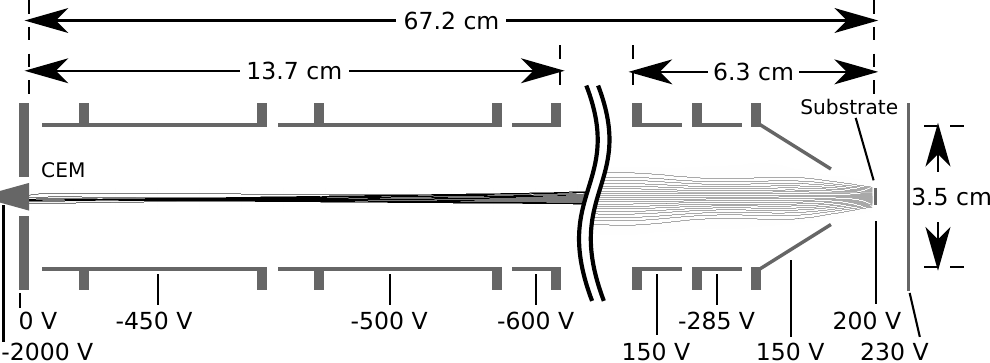}
	\caption{The ion optics of the TOF.  Ion flight paths simulated using SIMION are shown, in this case starting from a uniform distribution of ions with zero initial kinetic energy.  The ion flight paths are focused toward the CEM, accounting for the tighter beam at the detection end.  \label{fig:IonOptics}}
\end{figure*}

A time-of-flight (TOF) spectrometer (shown schematically in Figure~\ref{fig:IonOptics}) allows for mass analysis of the ionic species produced in LITD and RIS processes.  
As mentioned, non-Ba ions are produced by desorption from the substrate due to the LITD pulse, improperly shaped RIS pulses scraping the surface, non-resonant multi-photon processes and, in the case of the semiconducting substrate mounting technique, the Mo or Ta clips (described in more detail in Section~\ref{sec:Probe}) which may have a lower threshold for removal of ions.  A mass resolution on the order of $m/\Delta m \approx 100$ is sufficient to discriminate between the species of interest.  It is crucial that the ion transport efficiency of the TOF be very high to allow efficient Ba recovery.

To achieve this, ions are accelerated into the TOF by an electric field of $\sim60$~V/cm.  The first electrode in the TOF spectrometer is cone-shaped to allow for optical access while maintaining a strong electric field near the substrate.  Once the ions have been accelerated into the first lens triplet, they are collimated and drift through the vacuum tube.   At the end of the TOF path, ions are focused onto the $1$~cm diameter aperture of a channel electron multiplier (CEM)\cite{fn15} ion detector by a second lens triplet.  SIMION~\cite{Dahl:2000bu} simulations show a transport efficiency of $> 99$\% for ions from a thermal distribution appropriate for Ba desorbed as a neutral atom then ionized by the RIS process. 

\section{Substrate Mounting}
\label{sec:Probe}
The substrate must be positioned precisely in both the LXe cell and in the RIS chamber to ensure consistent deposition of Ba and laser desorption.  The substrate is mounted on the end of a $12.7$~mm-diameter, 85~cm long thin-walled stainless steel tube.  The vertical probe movement is actuated through a long bellows by a stepper motor outside the vacuum with a nominal vertical position accuracy of $3$~$\upmu$m.  Horizontal positioning is guaranteed by pairs of spring-loaded Vespel\cite{fn16} rollers that constrain the transverse motion of the probe and progressively engage the probe tube as it is lowered into the system. 

Substrates are mounted using two different methods depending upon the material.  The mounting system must not have protrusions that could adversely affect the electrostatics both for depositing Ba on the substrate and for the RIS process and also minimize the exposure of extraneous parts to the lasers impinging on the substrate.  The cleanliness of the substrate is ensured by \emph{in situ} cleaning systems, described in Sec.~\ref{sub:Cleaning}.  Bias voltages are applied through wiring running down the probe tube. 

\begin{figure}
	\includegraphics[width=3.1in, bb=233 305 379 486]{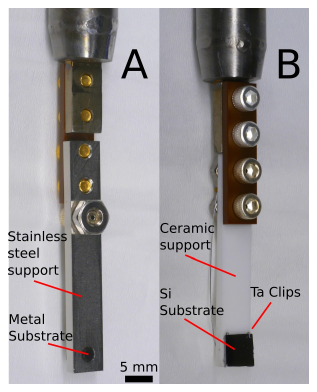}
	\caption{(Color Online) Two probe mounting systems: for refractory metal substrates (A) and for Si and other semiconductors (B).  The refractory metal substrate in A is held from behind by a stainless steel clip.  The substrate is isolated from the front plate.  The Si substrate in B is held by Mo or Ta wire clips.  \label{fig:ProbeTip}}
\end{figure}
Metallic substrates are held by a stainless steel clip (Figure~\ref{fig:ProbeTip}A), insulated from the surrounding stainless steel support by a MACOR\cite{fn17} insulator.  Silicon and silicon carbide substrates are held by Mo or Ta clips in turn supported by a MACOR support (Figure~\ref{fig:ProbeTip}B).  The clips serve as electrical contacts for resistive heating of the substrate for cleaning purposes, described further in Sec.~\ref{sub:Cleaning}.

\section{Data Acquisition}
\label{sec:DAQ}
Data acquisition from the TOF and laser systems is separate from the slow control system that regulates the LXe temperatures and pressures and actuates the probe.  TOF and laser data must be recorded at $10$~Hz and with sufficient bandwidth to separate pulses from multiple ions within each shot.  For each laser shot, a 250~MS/s FADC\cite{fn18} records data for $60~\upmu$s after the RIS laser pulse, long enough to record ions up to $300$~amu/e.  The CEM produces fast pulses amplified by a high-bandwidth preamp\cite{fn19}.  The same data acquisition system records the photodiode readings using 60~MS/s FADCs\cite{fn20}.  A LabVIEW program controls the shutters and a stage that can raster the desorption laser across the substrate. 

\section{System Performance}
\label{sec:Performance}
Successful Ba recovery depends upon several factors.  First, the RIS process must be tuned for efficient ionization of Ba.  Second, the substrate must be very clean to prevent Ba forming molecules with other adsorbed atoms (for example, oxygen) and to reduce desorbed ion signals in the TOF spectrum.  Finally, the TOF spectrometer must be calibrated to provide quantitative information on the species detected.  

\subsection{Basic Spectroscopy}
\label{sub:spectroscopy}
Efficient RIS requires that the atomic transitions are saturated.  Before designing the present system saturation conditions for both lasers were checked in a separate setup, using an atomic Ba beam.  This beam, produced by heating a Ba-coated tungsten wire inside a small volume with an aperture, emitted neutral Ba atoms at a far higher flux than the thermal desorption process described in Sec.~\ref{sub:Desorption}, simplifying the spectroscopy measurements. 

\begin{figure*}
	\includegraphics[scale=0.9]{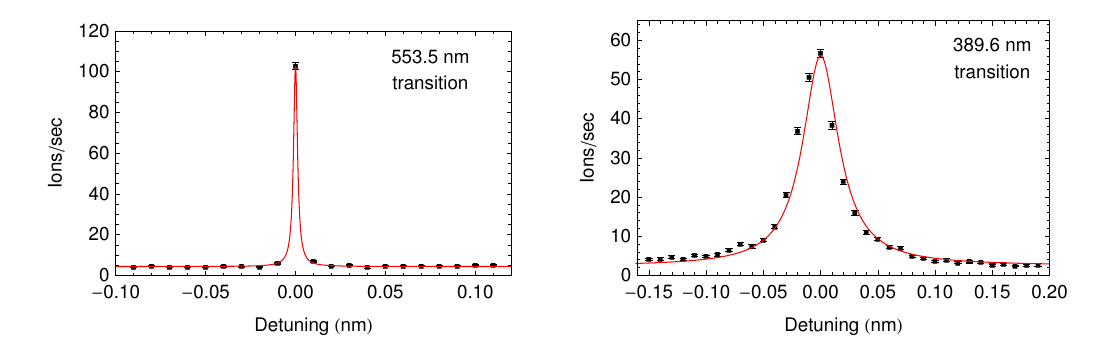}
	\caption{Ba$^+$ rates obtained by detuning the RIS lasers.  The left panel shows the 553.5~nm transition, the right shows the 389.6~nm transition.  A Lorentzian is fit to the 553.5~nm transition data.  A Fano profile (which accounts for interference between the atomic state and the continuum) is fit to the 389.6~nm transition.  Power and Doppler broadening are limited.  Errors shown are statistical. \label{fig:Detuning}}
\end{figure*}

The two RIS lasers were detuned in turn to obtain the spectra in Figure~\ref{fig:Detuning}. While scanning the 553.5 nm laser, the 389.6~nm laser power was maintained at 140~$\upmu$J/pulse with the laser on resonance. The scanning 553.5~nm laser was kept at 100~nJ/pulse and the resulting curve was fit to a Lorentzian ($\chi^2/ndf = 38/17$).  Errors on the data are statistical-only, possibly leading to large $\chi^2/ndf$ values.  The 389.6~nm laser was detuned while maintaining the 553.5~nm laser on resonance at 2.3~mJ/pulse.  The lineshape of this transition to an autoionizing state is described by a Fano profile\cite{Fano:1961ha}, which accounts for the interference between the atomic state and the continuum. In this case, $\chi^2/ndf = 194/30$ (still using statistical errors only).  

\begin{figure*}
	\includegraphics[scale=0.9]{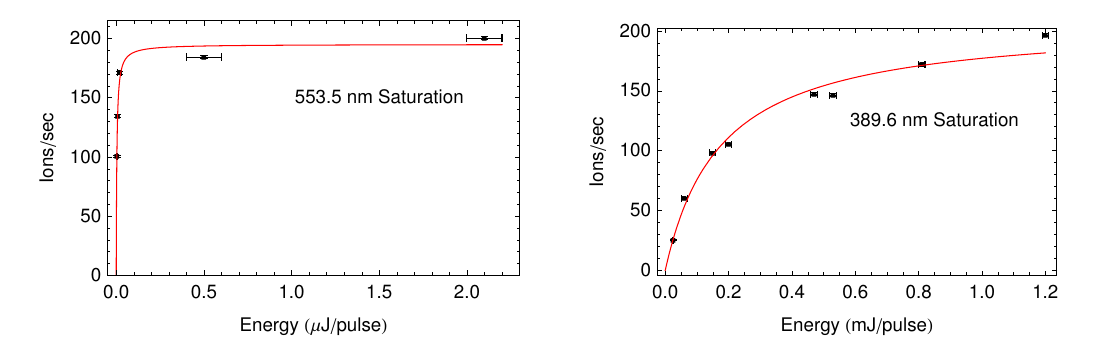}
	\caption{Saturation curves for the RIS lasers.  The saturation curves do not fit exactly because the saturation curve assumes a uniform laser profile and does not account for the ionization in the tails of the Gaussian beam.  These data were produced with a 2~mm $\sigma$ Gaussian beam.  Errors shown are statistical.  \label{fig:Saturation}}
\end{figure*}

Saturation curves are then obtained with both lasers on resonance while scanning the energy of each laser separately, as shown in Figure~\ref{fig:Saturation}.  The data are fit reasonably well by the model in reference~\cite{Metcalf:1999vv}, although $\chi^2/ndf$ values are large, presumably because systematic uncertainties are not taken into account.   

\subsection{Substrate Cleaning}
\label{sub:Cleaning}
Substrate preparation is crucial to clean adsorption and desorption of Ba.  We have used a number of different techniques to clean the substrate, depending upon the substrate material.  

Polished polycrystalline metal samples of Ta, W, and Ni were cleaned by electron bombardment heating to $1000^\circ$~C.  Temperatures were measured with a fiberoptic pyrometer\cite{fn21}.

Metallic substrates are also cleaned by low-energy sputtering using Xe ions \cite{Musket:1982go}.  In our system this is done by producing a plasma at 1~Torr pressure in the RIS chamber. The substrate, acting as a cathode, and a sharp anode biased at +1230~V with respect to the substrate create the plasma.  

Semiconductor substrates can be cleaned by resistive heating \cite{Musket:1982go}, applying the current through the Ta or Mo clips also used as a mechanical support.  These substrates can be heated to $1400^\circ$~C, although a typical cleaning cycle degasses the substrate at $700^\circ$~C for hours then removes the remaining oxide layer by ramping to $1200^\circ$~C for less than one minute.  This procedure is described to leave an atomically clean substrate \cite{Hu:2000eb}, although a base pressure of $7.5\times10^{-9}$~Torr leads to adsorbed layers of background gases in a few minutes.

\begin{figure}
	\includegraphics[scale=1, bb=4 2 238 162]{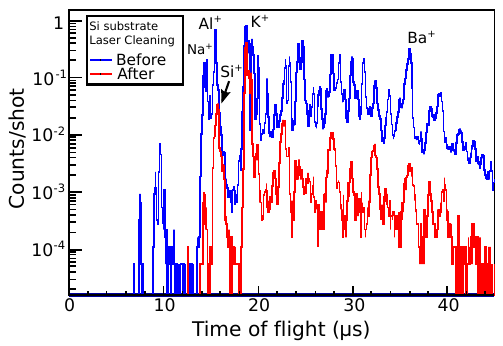}
	\caption{(Color online) Rastering the IR laser across the substrate is an effective cleaning technique.  This TOF spectrum shows the reduction in counts at all times of flight after rastering across the Si substrate with 1200~pulses per location at 1.2~mJ/pulse. \label{fig:LaserCleaning}}
\end{figure}

Finally, rastering the desorption laser across the substrate with $>1$~mJ/pulse also provides an effective cleaning scheme. The effect of such an IR laser scan is shown in the TOF spectrum in Figure~\ref{fig:LaserCleaning}.  

\subsection{Time-of-flight spectrometer calibration}
\label{sub:tofcalibration}
Calibration of the TOF spectrometer is necessary for effective identification of Ba.  Since several masses are required, ions generated in the LITD process are used.  From such a calibration, using the prominent Na$^+$ and K$^+$ peaks visible even at moderate pulse energies, the position of the Ba$^+$ peaks is inferred to be at $37\pm0.5~\upmu$s in the RIS spectrum, having accounted for the $1~\upmu$s RIS pulse delay and the fact that ionization due to RIS creates ions at slightly lower potentials, leading to a further delay of $\sim0.3~\upmu$s.  This position of the Ba$^+$ RIS peak is roughly consistent with the SIMION predicted TOF of $38.5~\upmu$s. 

The mass resolution near the Ba peak is found to be $m/\Delta m \approx 80$ by analyzing the shape of the $^{133}$Cs peak that is a common background.  While the TOF resolution is expected to depend upon the details of the ionization process, the mass difference between Ba isotopes of interest and the dominant background of $^{133}$Cs is sufficiently large to ensure proper separation even for resolutions somewhat worse than measured.

\begin{figure}
	\includegraphics[scale=1, bb=2 3 226 170]{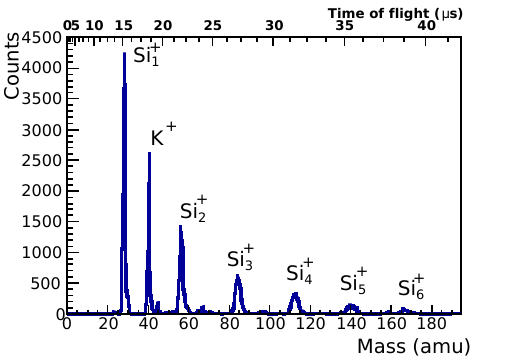}
	\caption{Higher laser powers desorb singly-charged Si$_n^+$ clusters.  The peak at 39~amu is potassium, a known surface contaminant. \label{fig:SiPeaks}}
\end{figure}

High desorption laser intensities remove singly ionized clusters of Si, up to Si$_6^+$ as shown in Figure~\ref{fig:SiPeaks}, providing a cross-check of the TOF calibration.  As expected, the rate of Si$^+_n$ clusters is a steep function of the desorption laser intensity and no Si$^+_n$ clusters are observed at LITD intensities used for the work with Ba.

\subsection{Detection of Barium from vacuum}
\label{sub:barium}
\begin{figure}
	\includegraphics[width=3.37in, bb=4 3 226 272]{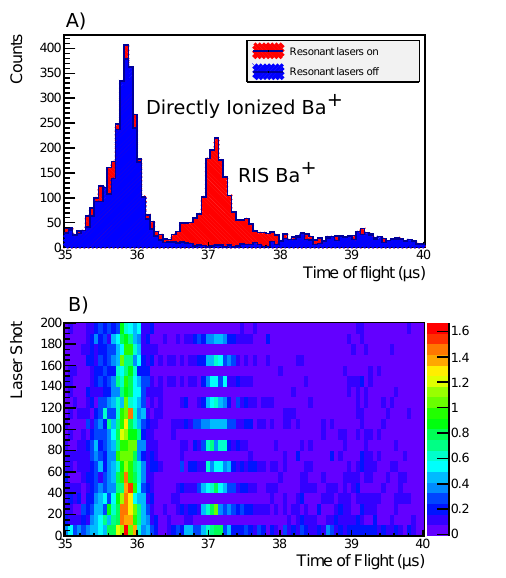}
	\caption{A typical TOF spectrum from Si~(100) with a strong Ba peak due to surface contamination of Ba.  In panel {\bf A}, the red hashed histogram shows data with the RIS lasers, while the blue histogram shows data without RIS lasers. The large excess in the red histogram at 37.1~$\upmu$s is desorbed and resonantly ionized Ba, while the peak at 35.8~$\upmu$s is Ba$^+$ desorbed by the desorption laser.  The resonantly ionized Ba peak is delayed relative to the directly ionized peak due to the $1~\upmu$s delay of the RIS lasers and different electrostatic conditions at the beginning of the TOF.  In panel {\bf B}, the change over 200 laser shots is shown.  The $y$-axis shows the laser pulse number up to 200 laser shots, with RIS lasers alternating on and off every 10~shots. \label{fig:ROITOF}}
\end{figure}
The combination of RIS and the TOF mass spectrometer allows positive identification of Ba contamination on a Si~(100) substrate cleaved from a commercially available wafer\cite{fn22} and mounted on the end of the probe.  The substrate was cleaned by resistive heating to $700^\circ$~C for 1~hour then flashed to $1200^\circ$~C for 30~seconds.  This particular sample produces a TOF spectrum with a significant Ba peak (Fig.~\ref{fig:ROITOF}), along with Na$^+$ and K$^+$ peaks that provide further confirmation of the TOF calibration.

Both resonantly ionized (peak at 37.1~$\upmu$s) and Ba$^+$ desorbed as an ion (35.8~$\upmu$s) are present in the TOF spectrum shown in Fig.~\ref{fig:ROITOF}.  The resonantly ionized Ba$^+$ peak is delayed due to the $1~\upmu$s delay of the RIS lasers and different electrostatic conditions at the beginning of the TOF.  The blue histogram contatains ions generated by the IR desorption laser alone, while the red histogram also includes the RIS lasers.  Peaks that appear in the red histogram but not the blue are due to ions undergoing the proper LITD+RIS process.  Note that in the laser pulsing sequence (Fig.~\ref{fig:ROITOF}B) the RIS Ba$^+$ peaks appears and disappears as expected.  Atoms and molecules ionized by the desorption laser alone are backgrounds to the RIS process. Detuning the dye lasers also suppresses the RIS Ba peak, further confirming that this is due to the LITD+RIS process.

\begin{figure}
	\includegraphics[scale=0.9]{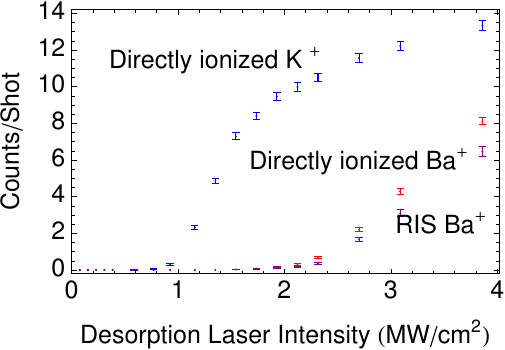}
	\caption{Desorption and ionization K$^+$ and Ba$^+$.  The desorption laser intensity is the average over the 7~ns laser pulse.  The desorption laser intensity was increased at a single location on a Si substrate.  Directly ionized K$^+$ has the lowest desorption threshold.  Directly ionized Ba$^+$ and Ba desorbed as a neutral atom then resonantly ionized appear with approximately the same desorption laser intensity threshold.  Uncertainties are statistical only.  RIS lasers are maintained at saturation.}
	\label{fig:Desorption}
\end{figure}

Variation of the desorption laser pulse energy allows measurement of the desorption thresholds for neutral atoms and ions.  Figure~\ref{fig:Desorption} shows ion yields for K$^+$ and Ba$^+$ as the desorption laser energy is increased at a single location on a Si substrate.  The threshold for desorbing and directly ionizing K is lower than that of Ba.  In the case of Ba both the direct ionization and LITD+RIS cases can be observed, maintaining constant the RIS pulse energies. The intensity threshold for desorption in these two processes appears to be approximately the same.  The rates and laser intensity thresholds for each of these processes vary with the substrate materials and preparations.

The study of substrate materials, their preparation and the optimal setup of the laser pulses to optimize the yield from the selective LITD+RIS process while minimizing the yield from direct ionization and other phenomena is the goal of the R\&D being performed with this apparatus. 

\section{Conclusions}
\label{sec:Conclusions}
An apparatus to develop Ba extraction and tagging from LXe has been developed.  The system is designed to insert a substrate into a LXe cell, load Ba atoms from a source and remove the substrate through a load-lock mechanism, allowing for laser desorption, resonant ionization and analysis of the ions recovered in a time of flight spectrometer.  Tests of the system in vacuum have shown detection of Ba from a substrate.  We identify Ba adsorbed onto the substrate through LITD and RIS, confirming the presence of Ba using a TOF mass spectrometer.  

Reducing backgrounds and increasing Ba recovery efficiency are important to the development of Ba tagging technology for the nEXO detector.  This setup allows development of cleaning techniques necessary for removal of the Ba contamination, testing substrates for appropriate adsorption and measurement of the efficiency of tagging with small numbers of ions.

\section{Acknowledgments}
This work is supported by the National Science Foundation, award No. PHY-1132382-001. We thank R.~Conley (SLAC), K.~Merkle and the Stanford Physics Machine shop for their help in constructing of the apparatus.  We thank H.~Manoharan (Stanford), J.~Schwede (Stanford) and P.~Vogel (Caltech) for many useful discussions.


\begin{thebibliography}{10}

\bibitem{Avignone:2008eo}
F.~Avignone, S.~R. Elliott, and J.~Engel,
\newblock Rev. Mod. Phys. {\bf 80}, 481 (2008).

\bibitem{Albert:2014ch}
J.~B. Albert {\em et~al.},
\newblock Nature {\bf 510}, 229 (2014).

\bibitem{Gando:2013wx}
A.~Gando {\em et~al.},
\newblock Phys. Rev. Lett. {\bf 110}, 062502 (2013).

\bibitem{Fancher:1979wf}
D.~Fancher {\em et~al.},
\newblock Nucl. Instr. Meth. Phys. {\bf 161}, 383 (1979).

\bibitem{Auger:2012up}
M.~Auger {\em et~al.},
\newblock J. Inst. {\bf 7}, P05010 (2012).

\bibitem{moe_detection_1991}
M.~K. Moe,
\newblock Phys. Rev. C {\bf 44}, R931 (1991).

\bibitem{hurst_resonance_1979}
G.~S. Hurst, M.~G. Payne, S.~D. Kramer, and J.~P. Young,
\newblock Rev. Mod. Phys. {\bf 51}, 767 (1979).

\bibitem{Green:2007vq}
M.~Green {\em et~al.},
\newblock Phys. Rev. A {\bf 76}, 023404 (2007).

\bibitem{montero_diez_simple_2010}
M.~Montero~D{\'\i}ez {\em et~al.},
\newblock Rev. Sci. Inst. {\bf 81}, 113301 (2010).

\bibitem{savard_radioactive_2008}
G.~Savard {\em et~al.},
\newblock Nucl. Instr. Meth. Phys. Res. B {\bf 266}, 4086 (2008).

\bibitem{fn1}
National Instruments Ethernet RIO NI-9148.

\bibitem{fn2}
SAES Pure Gas, Inc. model HP400-903F.

\bibitem{fn3}
MKS Baratrons, part numbers 627D53TBC1B and 627DU2TBE1B.

\bibitem{fn4}
MKS Baratron, Part Number 226A13TCDCDFB2A1.

\bibitem{fn5}
Pfeiffer Vacuum Full Range Gauges.

\bibitem{fn6}
Isotope Products, Valencia, CA.

\bibitem{Ziegler:2010wt}
J.~F. Ziegler, M.~D. Ziegler, and J.~P. Biersack,
\newblock Nucl. Instr. Meth. Phys. Res. B {\bf 268}, 1818 (2010).

\bibitem{England:C6UzhYbB}
T.~England and B.~Rider,
\newblock \emph{Evaluation and compilation of fission product yields 1993},
  (1995).

\bibitem{nndc}
National Nuclear Data Center, information extracted from the Chart of Nuclides
  database, http://www.nndc.bnl.gov/chart/. (June 2014).

\bibitem{Burgess:1986gy}
D.~Burgess, Jr, P.~C. Stair, and E.~Weitz,
\newblock J. Vac. Sci. Technol., A {\bf 4}, 1362 (1986).

\bibitem{fn7}
Minilite II, Continuum Lasers, $7$~ns pulse width.

\bibitem{fn8}
Thorlabs model FDS1010.

\bibitem{fn9}
Newport Power Meter Model 818P-001-12.

\bibitem{fn10}
Spectra Physics INDI-HG.

\bibitem{Willke:1993wt}
B.~Willke and M.~Kock,
\newblock J. Phys. B {\bf 26}, 1129 (1993).

\bibitem{Camus:1983vv}
P.~Camus, M.~Dieulin, A.~El~Himdy, and M.~Aymar,
\newblock Phys. Scripta {\bf 27}, 125 (1983).

\bibitem{fn11}
Sirah Cobra Stretch, using Pyrromethene 580 dye.

\bibitem{fn12}
Sirah Cobra Stretch with a mixture of Exalite~389 and Exalite~398 dye.

\bibitem{fn13}
Angstrom HighFinesse.

\bibitem{fn14}
National Instruments PXI-5105.

\bibitem{fn15}
DeTech model 2405.

\bibitem{Dahl:2000bu}
D.~A. Dahl,
\newblock Int. J. Mass Spectrom. {\bf 200}, 3 (2000).

\bibitem{fn16}
Vespel is a trademark of DuPont.

\bibitem{fn17}
MACOR is a trademark of Corning, Inc.

\bibitem{fn18}
National Instruments PXI-5114.

\bibitem{fn19}
Ortec VT120.

\bibitem{fn20}
National Instruments PXI-5105.

\bibitem{Fano:1961ha}
U.~Fano,
\newblock Phys. Rev. {\bf 124}, 1866 (1961).

\bibitem{Metcalf:1999vv}
H.~J. Metcalf and P.~van~der Straten,
\newblock {\em {Laser Cooling and Trapping}} (Springer, 1999).

\bibitem{fn21}
Omega model iR2P-600-53-C4EI.

\bibitem{Musket:1982go}
R.~G. Musket, W.~McLean, C.~A. Colmenares, D.~M. Makowiecki, and W.~J.
  Siekhaus,
\newblock Appl. Surf. Sci. {\bf 10}, 143 (1982).

\bibitem{Hu:2000eb}
X.~Hu {\em et~al.},
\newblock Surf. Sci. {\bf 445}, 256 (2000).

\bibitem{fn22}
Acquired from UniversityWafers.com, P/B, $0.1-1.0~\Omega{\rm cm}$, SSP.

\end{thebibliography}
\bibstyle{h-physrev}

\end{document}